\documentclass[twocolumn,preprintnumbers,amsmath,amssymb,pra]{revtex4}

\usepackage{graphicx}
\usepackage{dcolumn}
\usepackage{bm}
\usepackage{amssymb}
\usepackage{epstopdf}
\usepackage{color}
\usepackage{dsfont}
\usepackage{float}

\begin{document}

\title{Semiconductor quantum well irradiated by a two-mode electromagnetic field as a terahertz emitter}

\author{S. Mandal$^{1}$}
\author{T. C. H. Liew$^1$}
\author{O. V. Kibis$^{2,1}$}\email{Oleg.Kibis(c)nstu.ru}
\affiliation{$^1$Division of Physics and Applied Physics, School
of Physical and Mathematical Sciences, Nanyang Technological
University, Singapore 637371, Singapore}
\affiliation{${^2}$Department of Applied and Theoretical Physics,
Novosibirsk State Technical University, Karl Marx Avenue 20,
Novosibirsk 630073, Russia}

\begin{abstract}
We study theoretically the nonlinear optical properties of a
semiconductor quantum well (QW) irradiated by a two-mode
electromagnetic wave consisting of a strong resonant dressing
field and a weak off-resonant driving field. In the considered
strongly coupled electron-field system, the dressing field opens
dynamic Stark gaps in the electron energy spectrum of the QW,
whereas the driving field induces electron oscillations in the QW
plane. Since the gapped electron spectrum restricts the amplitude
of the oscillations, the emission of a frequency comb from the QW
appears. Therefore, the doubly-driven QW operates as a nonlinear
optical element which can be used, particularly, for optically
controlled generation of terahertz radiation.
\end{abstract}

\maketitle

\section{Introduction}
In recent time, lasers and microwave techniques are actively
exploited to control the nonlinear optoelectronic properties of
various systems with a strong high-frequency electromagnetic field
(dressing field). Since the field is intense, the mixing of
electron-field states results in a strongly coupled light-matter
object known as an electron dressed by the field (dressed
electron)~\cite{Scully_book,Cohen-Tannoudji_book}. The physical
properties of dressed electrons can be described within the
conventional Floquet theory of periodically driven quantum systems
~\cite{Hanggi_98,Kohler_2005,Bukov_2015,Holthaus_2016} and have
been studied for various nanostructures, including semiconductor
quantum wells
\cite{Wagner_10,Kibis_2012,Teich_13,Dini_16,Morina_15}, quantum
dots~\cite{Savenko_2012,Kryuchkyan_2017}, quantum rings
\cite{Kibis_2011,Sigurdsson_14,Koshelev_15,Joibari_14}, graphene
\cite{Glazov_14,Perez_14,Syzranov_13,Oka_09,Kibis_17}, etc. Among
a variety of nonlinear optical phenomena induced by a dressing
field, the opening of dynamic Stark gaps in electron energy
spectra of various condensed-matter structures should be noted
especially (see, e.g.,
Refs.~\onlinecite{Goreslavskii_1969,Kibis_2011,Kibis_2012,Vu_2004,Wang_2013}).
In the present research, we demonstrate theoretically that this
nonlinear effect results in the emission of a terahertz frequency
comb from a semiconductor quantum well (QW) irradiated by a
two-mode electromagnetic field.

\section{Model}
Let us consider a two-dimensional electron gas (2DEG) in a
semiconductor QW irradiated by a linearly polarized two-mode
electromagnetic wave, which propagates along the $z$ axis and
consists of a strong high-frequency dressing field,
$\widetilde{E}\cos\omega_0 t$, with the electric field amplitude,
$\widetilde{E}$, and the frequency, $\omega_0$, and a relatively
weak and low-frequency driving field, $E\cos\omega t$, with the
electric field amplitude, $E$, and the frequency, $\omega$ (see
Fig.~1a). In what follows, we will assume that the dressing field
frequency, $\omega_0$, satisfies the condition
$\hbar\omega_0>\varepsilon_g$, where $\varepsilon_g$ is the band
gap of the QW. In this case, the dressing field strongly mixes
electron states from the valence and conduction bands and,
therefore, substantially modifies their energy spectrum due to the
dynamic Stark effect. Particularly, the dressing field opens
energy gaps, $\Delta\varepsilon$, within the bands in the resonant
points of the Brillouin zone, $k_0$, where the photon energy,
$\hbar\omega_0$, is equal to the energy interval between the
bands~\cite{Goreslavskii_1969}. This gap opening is shown
schematically in Figs.~1b and 1c, where the first valence and
conduction subbands of the QW are plotted in the Schr\"odinger and
interaction pictures, correspondingly. For definiteness, we will
restrict the following analysis by the Schr\"odinger picture. In
the most general case, the energy spectrum of dressed electrons
arising from any two bands mixed by the dressing field,
$\varepsilon^+(\mathbf{k})$ and $\varepsilon^-(\mathbf{k})$, can
be written in the form~\cite{Kibis_2012}
\begin{equation}\label{E}
\varepsilon(\mathbf{k})=\frac{\varepsilon^+(\mathbf{k})+\varepsilon^-(\mathbf{k})}{2}
\pm\frac{\hbar\omega_0}{2}
\pm\frac{\alpha(\mathbf{k})}{2|\alpha(\mathbf{k})|}
\sqrt{(\hbar\Omega_R)^2+\alpha^2(\mathbf{k})},
\end{equation}
where
$\alpha(\mathbf{k})=\varepsilon^+(\mathbf{k})-\varepsilon^-(\mathbf{k})-\hbar\omega_0$
is the resonance detuning, $\mathbf{k}$ is the electron wave
vector, $\Omega_R=d\widetilde{E}/\hbar$ is the Rabi frequency of
interband electron transitions, and $d$ is the interband dipole
moment. In the considered case of a semiconductor QW, these two
branches, $\varepsilon^+(\mathbf{k})$ and
$\varepsilon^-(\mathbf{k})$, should be treated as the first
electron subband,
$\varepsilon^+(\mathbf{k})=\hbar^2k^2/2m_e+\varepsilon_g/2$, and
the first hole subband,
$\varepsilon^-(\mathbf{k})=-\hbar^2k^2/2m_h-\varepsilon_g/2$,
where $m_e$ and $m_h$ are the electron and hole masses,
respectively, and $\mathbf{k}=(k_x,k_y)$ is the electron wave
vector in the QW plane. Since the dynamic Stark gaps,
$\Delta\varepsilon=\hbar\Omega_R$, break the dispersion curves
(\ref{E}) only at the border wave vector, $|\mathbf{k}|=k_0$, the
energy spectrum (\ref{E}) around the ground state of the
conduction band (see the red dot in Fig.~1b) is continuous and can
be written as
\begin{equation}\label{Ee}
\varepsilon_e(\mathbf{k})=\frac{\hbar^2k^2}{2m_-}
+\frac{\hbar\omega_0}{2}-\frac{1}{2}
\sqrt{(\Delta\varepsilon)^2+\left[\hbar\omega_0-\varepsilon_g-\frac{\hbar^2k^2}{2m_+}\right]^2},
\end{equation}
where $m_\pm=m_em_h/(m_e\pm m_h)$ is the reduced electron-hole
mass, the electron wave vector, $\mathbf{k}$, lies within the
range $|\mathbf{k}|\leq k_0$, and the border wave vector, $k_0$,
satisfies the condition
$\hbar\omega_0=\varepsilon_g+\varepsilon_0$ with the
characteristic border electron energy,
$\varepsilon_0=\hbar^2k_0^2/2m_+$. It should be stressed that the
dressing field amplitude, $\widetilde{E}$, is assumed to satisfy
the strong light-matter coupling condition, $\Omega_R\tau\gg1$,
where $\tau$ is the mean free time of conduction electrons in the
QW. In this regime, the interband absorption of the dressing field
is absent (see, e.g., Ref.~\cite{Kibis_2012}) and, therefore, the
dressing field does not generate electron-hole pairs.
\begin{figure}[h!]
\begin{center}
\includegraphics[width=1\linewidth]{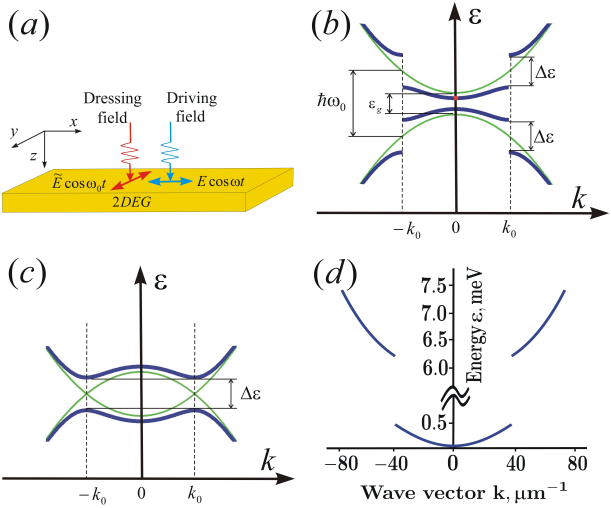}
\caption{Sketch of the system under consideration: (a)
Two-dimensional electron gas (2DEG) in a semiconductor quantum
well (QW) irradiated by a two-mode electromagnetic field
consisting of a dressing field with the electric field amplitude,
$\widetilde{E}$, and the frequency, $\omega_0$, and a driving
field with the electric field amplitude, $E$, and the frequency,
$\omega$; (b) Energy spectrum of the first conduction and valence
subbands of the QW with the band gap, $\varepsilon_g$, in the
presence of the dressing field (the blue lines) and without the
field (the green lines) in the Schr\"odinger picture. Due to the
gaps, $\Delta\varepsilon$, induced by the dressing field, the
driving field can move a conduction electron (the red dot) along
the dispersion curve only between the states $-k_0$ and $k_0$; (c)
Energy spectrum of the first conduction and valence subbands of
the QW in the presence of the dressing field (the blue lines) and
without the field (the green lines) plotted in the interaction
picture; (d) Energy spectrum of the first electron subband in a
GaAs-based QW irradiated by a dressing field with the frequency
$\hbar\omega_0=\varepsilon_g+1\,\mathrm{meV}$ and the amplitude
$\widetilde{E}=10^7\,\mathrm{V/m}$.} \label{fig:fig1}
\end{center}
\end{figure}
As to the driving field, its frequency is relatively low,
$\hbar\omega\ll\varepsilon_g$. Therefore, the off-resonant driving
field does not mix almost electronic states from the conduction
and valence bands and, correspondingly, does not change the energy
spectrum (\ref{E}) renormalized by the resonant dressing field.
Thus, the driving field results only to the electron transitions
between different states of the continuous spectrum (\ref{Ee}),
which can be described within the conventional semiclassical
approach as follows. First of all, it should be stressed that the
magnetic component of the driving field lies in the plane $(x,y)$.
Since the electron velocities in the QW, $\mathbf{v}=(v_x,v_y)$,
lie in the same plane, the Lorentz force caused by the magnetic
field is directed perpendicularly to the QW. Therefore, it cannot
affect electron dynamics in the QW plane. As a consequence, the
dynamics of dressed electrons in the QW depends only on the
electric component of the driving field, $\mathbf{E}=(E\cos\omega
t,\,0)$. Within the continuous energy spectrum (\ref{Ee}), the
electron dynamics can be described by the energy conservation law,
$e\mathbf{v}\mathbf{E}={\partial
\varepsilon_e(\mathbf{k})}/{\partial t}$, where the left part is
the energy received by a dressed electron from the driving field
per unit time. It follows from the exact solution of the
electron-photon Schr\"odinger problem~\cite{Kibis_2012} that the
averaged velocity of the dressed electron in the state with the
wave vector $\mathbf{k}$ is given by the expression
$\mathbf{v}(\mathbf{k})=(1/\hbar){\partial\varepsilon_e(\mathbf{k})}/{\partial\mathbf{k}}$,
which coincides formally with the well-known classical Hamilton
equation,
$\mathbf{v}(\mathbf{p})=\partial\varepsilon(\mathbf{p})/\partial\mathbf{p}$,
describing the velocity of a particle with an energy
$\varepsilon(\mathbf{p})$ and the generalized momentum
$\mathbf{p}=\hbar\mathbf{k}$. Substituting this electron velocity,
$\mathbf{v}(\mathbf{k})$, into the energy conservation law, we
arrive at the conventional dynamics equation,
\begin{equation}\label{k}
\hbar\frac{d\mathbf{k}}{dt}\Big|_{|\mathbf{k}|< k_0}=e\mathbf{E}.
\end{equation}
Since the driving field is substantially weaker than the dressing
field, a dressed electron cannot surmount the dynamic Stark gap,
$\Delta\varepsilon$, at the border wave vector, $|\mathbf{k}|=k_0$
(see Fig.~1b). Taking this energy barrier into account, one has to
complement the semiclassical dynamics equation (\ref{k}) with the
boundary condition,
\begin{equation}\label{k0}
\hbar\frac{d\mathbf{k}}{dt}\Big|_{|\mathbf{k}|=k_0}=\left\{\begin{array}{rl}
0,
&e\mathbf{k}\mathbf{E}>0\\
e\mathbf{E},&e\mathbf{k}\mathbf{E}\leq0
\end{array}\right..
\end{equation}
The three expressions (\ref{Ee}), (\ref{k}) and (\ref{k0}) give
the complete description of the single-electron dynamics in the
irradiated QW. For definiteness, let us apply them to consider an
electron which is initially in the ground state, $\mathbf{k}=0$
(see the red dot in Fig.~1b). In this case, the electron wave
vector depends on time, $t$, as a function
$\mathbf{k}(t)=(k_x(t),0)$, where
\begin{equation}\label{K}
k_x(t)=\left\{\begin{array}{rl} k_E\sin\omega t,
&|k_E\sin\omega t|<k_0\\
k_0,&k_E\sin\omega t\geq k_0\\
-k_0,&k_E\sin\omega t\leq -k_0
\end{array}\right..
\end{equation}
and $k_E=eE/\hbar\omega$ is the amplitude of electron oscillations
in $k$-space. The oscillating movement of the electron, which
arises from the sines in Eq.~(\ref{K}), should result in the
emission of electromagnetic waves from the QW. Restricting
analysis of the emission by the dipole approximation, we can write
the intensity of the emission at the frequency of the $n$-th
harmonic, $\omega_n=n\omega$, in the conventional form (see, e.g.,
Ref.~\cite{Landau_v2}),
\begin{equation}\label{I}
I_n=\frac{4e^2}{3c^3}|{\mathbf{a}}_n|^2,
\end{equation}
where $\mathbf{a}_n$ is the $n$-th coefficient of the Fourier
expansion of the electron acceleration, $\mathbf{a}=(a_x,0)$. It
follows from Eq.~(\ref{Ee}) that the electron acceleration,
${a}_x=(\dot{v}_x,0)$, can be written in the dimensionless form as
\begin{align}\label{Ac}
&\frac{a_x}{a_0}=\frac{\dot{k}_x}{\omega k_0}
\Bigg\{2\left(\frac{m_+}{m_-}\right)+2\Bigg(\left[\frac{{\Delta\varepsilon}}{\varepsilon_0}\right]^2+
\left[1-\frac{k^2}{k_0^2}\right]^2\Bigg)^{-\frac{3}{2}}\nonumber\\
&\times
\Bigg[1-\frac{k^2}{k_0^2}\Bigg]^2\frac{k_x^2}{k_0^2}+\left(\left[\frac{{\Delta\varepsilon}}{\varepsilon_0}\right]^2+
\left[1-\frac{k^2}{k_0^2}\right]^2\right)^{-\frac{1}{2}}\nonumber\\
&\times\Bigg[1-\frac{k^2}{k_0^2}-2\frac{k_x^2}{k_0^2}\Bigg]\Bigg\},
\end{align}
where $a_0=\hbar k_0\omega/2m_+$ is the characteristic electron
acceleration. The expressions (\ref{K})--(\ref{Ac}) define the
spectrum of electromagnetic emission from the QW, which is
discussed below. For definiteness, we will proceed with the
analysis of a GaAs-based QW ($\varepsilon_g=1.39$~eV,
$m_e=0.067\,m_0$, $m_h=0.47\,m_0$) which is a suitable system for
experimental observation of the discussed effects (see the
corresponding electron energy spectrum in Fig.~1d).

\section{Results and discussion}
The features of the electromagnetic emission from the QW originate
from the dynamic Stark gaps, $\Delta\varepsilon$, which take place
at the border electron wave vector, $|\mathbf{k}|=k_0$ (see
Fig.~1b). Physically, these gaps confine electron oscillations
induced by the driving field within the domain $-k_0\leq k_x\leq
k_0$ and, therefore, crucially affect the emission spectrum. It
follows from Eq.~(\ref{K}) that the electron wave vector (\ref{K})
depends on the time, $t$, as a purely harmonic function,
$k_x(t)=k_E\sin\omega t$, if $k_E\leq k_0$. As a consequence, the
electron acceleration, $\dot{a}_x\sim\dot{k}_x\sim\cos\omega t$,
is also a harmonic function in this case. This results in the
electromagnetic emission from the QW only with the frequency of
the driving field, $\omega$, if $k_E\leq k_0$. However, if the
amplitude, $k_E$, satisfies the condition $k_E>k_0$, the electron
oscillations (\ref{K}) are stopped in $k$-space at the border wave
vector, $k_0$. This leads to the ``clipped'' oscillations of both
the wave vector, $k_x(t)$, and the acceleration, $a_x(t)$, which
are plotted in Fig.~2a. The expansion of the clipped oscillations
of the acceleration, $a_x(t)$, into the Fourier series leads to
non-zero intensities (\ref{I}) with multiple frequencies,
$\omega_n=n\omega$. These intensities are plotted in Fig.~2b as
$I_n/I_0$, where $I_0=(e\hbar\omega k_0/m^+c)^2/3c$ is the
characteristic intensity. It should be noted that the Fourier
expansion of the acceleration, $a_x(t)$, pictured in Fig.~2a, has
only odd harmonics with $n=1,3,5,...$. Therefore, corresponding
odd harmonics of the intensity, $I_n$, appear  in the emission
spectrum plotted in Fig.~2b. As expected, only the harmonic with
$n=1$ is significant in Fig.~2b for low driving field amplitudes,
$E/E_0\ll1$, where $E_0=\hbar\omega k_0/e$ is the characteristic
field strength. As we keep on increasing the driving field
amplitude, $E$, other harmonics of the emission with $n>1$ come
into play in Fig~2b. As a consequence, a frequency comb of the
electromagnetic emission from the QW appears.

It follows from the aforesaid that the discussed mechanism of
high-frequency generation arises from the nonparabolic area of the
electron energy spectrum (\ref{Ee}), which takes place near the
border electron wave vector, $k_0$. A physically similar mechanism
exists also in semiconductor QWs with strongly nonparabolic
electron dispersion~\cite{Haljan_2003,Hurst_2016} and
superlattices~\cite{Hyart_2009}. However, in these systems the
nonparabolic region lies very far from the band edge. Therefore,
the driving field should be very strong for high-frequency
generation. In contrast to this case, the present effect arises
from the dynamic Stark gaps which can be induced by a dressing
field very close to the band edge (i.e., the border wave vector,
$k_0$, can be very small). As a consequence, the generation
appears for a relatively weak driving field. The high-frequency
generation in such a near-band-edge regime (which, particularly,
corresponds to low electron densities) is the main advantage of
the present mechanism. It should be noted that the minimal value
of the border electron wave vector, $k_0$, is restricted only by
the Heisenberg uncertainty relation. Namely, the condition
$k_0l\gg1$ should be satisfied, where $l$ is the mean free path of
electrons in the QW. Since the path, $l$, is macroscopically large
in state-of-the-art QWs, the border wave vector, $k_0$, can be
very small.
\begin{figure}[h!]
\includegraphics[width=0.8\linewidth]{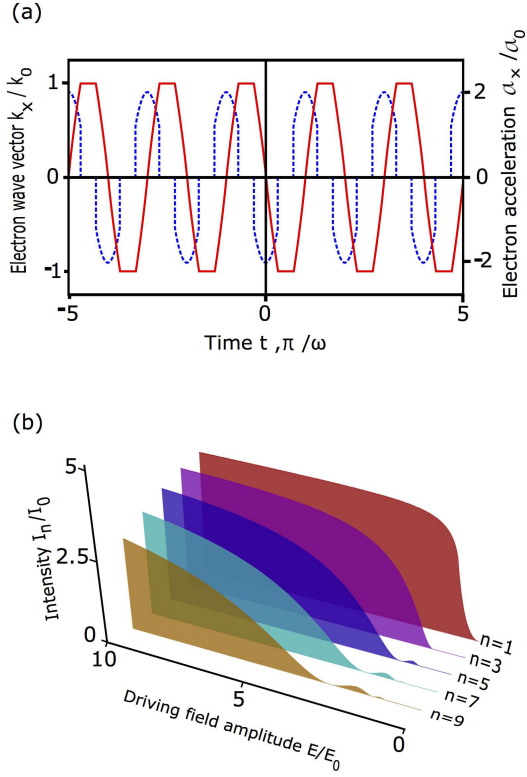}
\caption{Scheme of the single-electron electromagnetic emission
from the quantum well with the Stark gap
$\Delta\varepsilon/\varepsilon_0=50$: (a) Oscillation dynamics of
electron wave vector, $k_x/k_0$ (the dashed line), and electron
acceleration, $a_x/a_0$ (the solid line) for  the driving field
amplitude ${E}/E_0=1.14$; (b) Intensity of the electromagnetic
emission, $I_n$, with the different frequencies,
$\omega_n=n\omega$, as a function of the driving field amplitude,
$E$.} \label{fig:fig2}
\end{figure}
\begin{figure}[h!]
\includegraphics[width=0.7\linewidth]{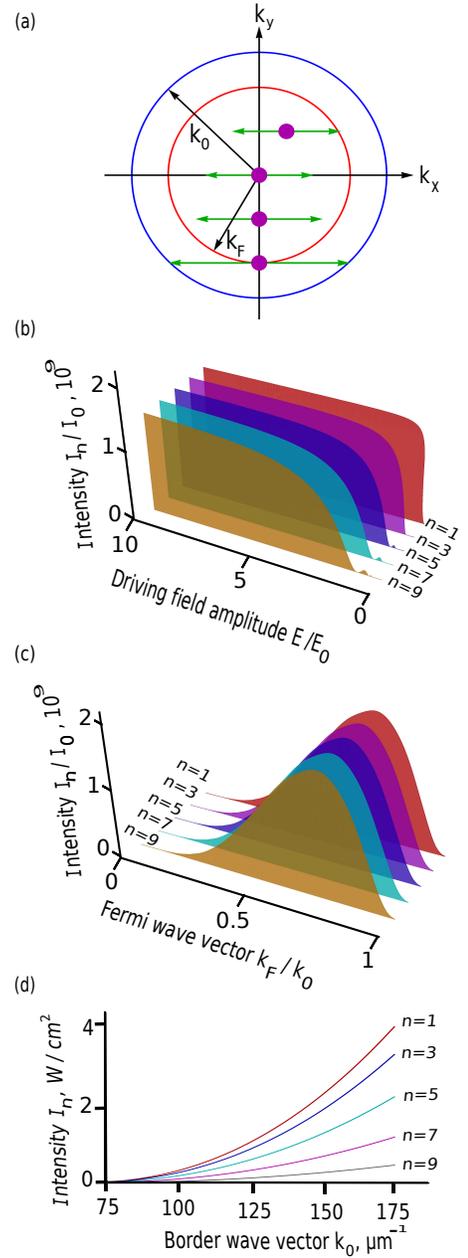}
\caption{Scheme of the multi-electron electromagnetic emission
from the quantum well: (a) Sketch of electron oscillations under
the driving field; (b) Intensity of the electromagnetic emission,
$I_n$, with the different frequencies, $\omega_n=n\omega$, as a
function of the driving field amplitude, $E$, for the Stark gap
$\Delta\varepsilon/\varepsilon_0=50$ and the Fermi wave vector
$k_F/k_0$=0.7; (c) Intensity of the electromagnetic emission,
$I_n$, as a function of the Fermi wave vector, $k_F$, for the the
Stark gap $\Delta\varepsilon/\varepsilon_0=50$ and the driving
field amplitude $E/E_0=9.8$; (d) Intensity of the electromagnetic
emission, $I_n$, from a GaAs-based QW with the planar dimensions
$100\times100\,\mu$m as a function of the border wave vector,
$k_0$, for the the Stark gap $\Delta\varepsilon=5$ meV, the
characteristic wave vector difference, $k_0-k_F=25$
$\mu\mathrm{m}^{-1}$, and the driving field with the frequency
$\omega=10^{12}$~rad/s and the irradiation intensity
$I=1$~kW/cm$^2$.} \label{fig:fig4}
\end{figure}

To go from the considered single-electron model to the case of
2DEG in real semiconductor QWs (see, e.g., Ref.~\cite{Haug_book}),
we have to make several assumptions. To neglect excitonic effects
and intersubband electron dynamics, let us assume that the field
frequencies are far from both the excitonic frequency and the
frequencies corresponding to intersubband electron transitions. To
avoid the destructive effect of scattering processes on the
electron oscillations, we will assume that the driving field
frequency, $\omega$, is high enough to meet the condition
$\omega\tau\gg1$, where $\tau$ is the mean free time of conduction
electrons restricted by the electron scattering with phonons and
impurities. In modern QWs, this condition can be easily satisfied
for driving field frequencies, $\omega$, starting from the
sub-terahertz range. It should be noted that the simple model that
only accounts for intraband acceleration and ignores all
scattering is already sufficient to successfully describe the
high-harmonic generation in various nonlinear systems (see, e.g.,
Refs.~\onlinecite{Mucke_2011,Luu_2015}). Particularly, more
advanced studies (e.g., based on the Bloch equations
technique~\cite{Golde_2008,Shubert_2014}) do not dramatically
change the results from the simple model. It should be noted also
that the confinement potential of QW does not influence on the
discussed effect since the polarization vector of the field lies
in the plane of QW (see Fig. 1a). As to temperature, it should be
less than the Stark gap to avoid the thermal excitation of
electrons over the gap. In typical semiconductor materials, the
gap is of meV scale for the dressing field amplitudes
$\widetilde{E}\sim10^7\,\mathrm{V/m}$~\cite{Goreslavskii_1969}.

To extend the single-electron model to the case of a
multi-electron system in the QW, let us consider a degenerate 2DEG
with the Fermi wave vector $k_F<k_0$, which fills electron states
(\ref{Ee}) at zero temperature. Physically, the electromagnetic
emission from the 2DEG can be described within the same model
(\ref{E})--(\ref{Ac}), but complemented with the Pauli principle.
The difference between the single-electron and the multi-electron
models is illustrated schematically in Fig.~3a. Namely, the
distribution of the 2DEG in the $k$-space can be described by the
Fermi circle (see the red line in Fig.~3a). Under the driving
field, $E$, all electrons oscillate along the $x$ axis with
different amplitudes (see the green arrows in Fig.~3a), which are
restricted by the Pauli principle and strongly depend on initial
positions of the electrons in $k$-space (see the violet dots in
Fig.~3a). As a result, in contrast to the single-electron model,
not all electrons can reach the gap at the border wave vector
$k_0$ (see the blue circle in Fig.~3a). Taking the Pauli principle
into account, Eq.~(\ref{K}) for an electron from the 2DEG should
be rewritten as
\begin{equation}\label{K10}
k_x(t)=\left\{\begin{array}{rl} k_x(0)+k_E\sin\omega t,
&k_x^{-}<k_E\sin\omega t<k_x^{+}\\
k_x(0)+k_x^+,&k_E\sin\omega t\geq k_x^+\\
k_x(0)+k_x^-,&k_E\sin\omega t\leq k_x^-
\end{array}\right.,
\end{equation}
where
$k_x^{\pm}=\pm\left(\sqrt{k_0^2-k_y^2}-\sqrt{k_F^2-k_y^2}\right)$
and $k_{x}^2(0)+k_{y}^2\leq k_F^2$. Substituting Eq.~(\ref{K10})
into Eq.~(\ref{Ac}) and performing summation of the electron
acceleration (\ref{Ac}) over filled electron states, we arrive at
the total acceleration of the multi-electron system, $\mathbf{a}$.
Assuming the QW size to be less than the wavelength of the
electromagnetic emission, the emission intensity at the frequency
$\omega_n=n\omega$ can be described by Eq.~(\ref{I}). As a result,
we arrive at the intensities of different harmonics, $I_n$, which
are plotted in Figs.~3b--3c. Due to the contribution from many
electrons, the radiation emitted from the 2DEG is much stronger
than the radiation emitted by a single electron (see Figs.~2b and
3b). It should be noted that the Fermi energy in GaAs-based QWs
can be easily controlled by the gate voltage without doping. If we
increase the Fermi wave vector, $k_F$, the intensities, $I_n$,
also increase since the total number of electrons increase (see
Fig.~3c). However, the intensities, $I_n$, start to decrease if
the Fermi wave vector, $k_F$, comes very close to $k_0$ (see
Fig.~3c). Indeed, in this case a large number of electrons have
insufficient empty states in which to oscillate. It should be
noted also that the Fermi energies in Figs.~3b and 3c lie in the
sub-meV range. This corresponds to very low electron densities,
such that electron-electron processes do not affect our results
and can be neglected. Particularly, the characteristic energy of
electron-electron interaction is also of sub-meV scale and,
therefore, sufficiently less than the Stark gap induced by the
dressing field. The emission power associated with each harmonic,
$I_n$, as a function of the border wave vector, $k_0$, is plotted
in Fig.~3d. Tuning the photon energy of the dressing field,
$\hbar\omega_0$, we can easily control the border wave vector,
$k_0$. If the difference of the characteristic wave vectors,
$\Delta k=k_0-k_F$, is fixed, the increasing of the border wave
vector, $k_0$, results in increasing the total number of
electrons. This leads to an enhancement of the output power (see
Fig.~3d). But when the border wave vector, $k_0$, becomes too
large, many electrons cannot reach the Stark gaps. As a result,
only the first few harmonics gain power and the power of higher
harmonics starts to decrease.

The discussed mechanism of electromagnetic emission from the QW
looks most attractive for the generation of frequency combs in the
terahertz range. Indeed, applying a driving field with the
sub-terahertz frequency, $\omega=10^{12}$~rad/s, we can
effectively produce the terahertz emission with the frequencies
$\omega_n=n\omega$ ($n=3,5,7,...$) and the intensities of a few
W/cm$^2$ (see Fig.~3d). It should be noted that frequency combs
are highly sought for making optical clocks, for measuring
fundamental constants accurately~\cite{Rosenband_2008}, for high
resolution molecular spectroscopy~\cite{Adler_2010}, etc.
Traditionally, frequency combs are generated using pulsed
lasers~\cite{Ferguson_2002} and quantum cascade
lasers~\cite{Burghoff_2014,Bartalini_2014}. In contrast to these
systems, the present comb  emitter is very compact. This is
important since the search for compact reliable sources of
terahertz emission is one of the exciting fields of modern applied
physics. In particular, terahertz scattering processes between
polariton branches~\cite{Kavokin_2010,Savenko_2011,Huppert_2014},
intersubband polaritons~\cite{DeLiberato_2013},
dipolaritons~\cite{Kyriienko_2017}, and different orbital
states~\cite{Kavokin_2012,Pervishko_2013} have been considered.
Transitions between direct and indirect exciton states, where
nonlinearity allows continuous oscillation and terahertz output as
a result of parametric instability have also been
studied~\cite{Kyriienko_2013,Kristinsson_2013}. Moreover, bosonic
cascade lasers in which multiple terahertz transitions can be
chained have been
proposed~\cite{Liew_2013,Liew_2016,Kaliteevski_2014}. It should be
noted that terahertz frequency combs from exciton-polariton
systems have also been proposed, but their expected efficiency is
largely unknown~\cite{Rayanov_2015}. It follows from the aforesaid
that the present theory describing the efficient terahertz
emission from the strongly coupled electron-light system in a QW
fits well current tendencies in terahertz physics.

In conclusion, we have demonstrated theoretically that a
semiconductor QW irradiated by a two-mode electromagnetic field
can serve as an effective source of a terahertz frequency comb.
Physically, the discussed nonlinear optical effect arises from the
dynamic Stark gaps in the electron energy spectrum of a QW, which
strongly modify the electron dynamics in the QW and results in the
terahertz emission. Since the parameters affecting the terahertz
emission from the QW can be tuned by irradiation, the present
theory paves a way to optically controlled QW-based terahertz
emitters.

The work was partially supported by the Ministry of Education of
Singapore AcRF Tier 2 (Project No. MOE2015-T2-1-055), Russian
Foundation for Basic Research (Project No 17-02-00053) and
Ministry of Education and Science of Russian Federation (Project
No 3.4573.2017/6.7).

\end{document}